\documentclass[%
  reprint,%
  amssymb, amsmath,%
   aip,jmp,%
    nofootinbib
]{revtex4-2}

\usepackage{bm}%
\usepackage[colorlinks=true,linkcolor=blue,urlcolor=blue,citecolor=blue]{hyperref}%
\usepackage{orcidlink}

\begin{document}

\title{Pre-Supernova (Anti)Neutrino Emission Due to  Weak-Interaction Reactions with Hot Nuclei}
\thanks{Published in \href{https://doi.org/10.3390/particles8040084}{Particles {\bf 8}, 84 (2025)}} 

\author{Alan A. Dzhioev\orcidlink{0000-0003-0642-1079}}
\email{dzhioev@theor.jinr.ru}
\affiliation{Bogoliubov Laboratory of Theoretical Physics, Joint Institute for Nuclear Research, 141980 Dubna, Russia}

\author{Andrey V. Yudin\orcidlink{0000-0002-0986-4257}}
\email{yudin@itep.ru}
\affiliation{National Research Center “Kurchatov Institute”, 123182 Moscow, Russia}

\author{Natalia V. Dunina-Barkovskaya}
\email{dunina@itep.ru}
\affiliation{National Research Center “Kurchatov Institute”, 123182 Moscow, Russia}

\author{Andrey I. Vdovin}
\email{vdovin@theor.jinr.ru} 
\affiliation{Bogoliubov Laboratory of Theoretical Physics, Joint Institute for Nuclear Research, 141980 Dubna, Russia}

\begin{abstract}
Reliable predictions of (anti)neutrino spectra and luminosities are essential for assessing the feasibility of detecting pre-supernova neutrinos. Using the stellar evolution code MESA, we calculate the (anti)neutrino spectra and luminosities under realistic conditions of temperature, density,  and electron fraction.  Our study includes (anti)neutrinos produced by both thermal processes and nuclear weak-interaction reactions.   By comparing the results of the thermal quasiparticle random-phase approximation   with the standard technique  based on the effective $Q$-value method, we investigate how thermal effects influence the spectra and luminosities of emitted (anti)neutrinos. Our findings show that a thermodynamically consistent treatment of Gamow--Teller transitions in hot nuclei enhances both the energy luminosity and the average energies of the emitted (anti)neutrinos.
\end{abstract}


\maketitle

\section{Introduction}\label{intro}

It is now widely recognized that in stellar interiors characterized by high temperatures and densities, the~emission of neutrinos and antineutrinos plays a significant role in energy loss, entropy removal from the stellar core, and~the acceleration of stellar evolution~\cite{Woosley_RevModPhys74,Janka_PhysRep442,Balasi_PPNP85}. The~detection of neutrino bursts from SN~1987A~\cite{Hirata1987PhRvL,Bionta1987PhRvL,Alekseev1987JETPL} significantly advanced our understanding of supernova mechanisms and stimulated the research in neutrino astrophysics. With~substantial advancements in neutrino detection technologies, the~potential for observing neutrino signals from new astrophysical sources has been extensively explored in recent decades. It is anticipated that neutrinos  and antineutrinos emitted by stars in the pre-supernova phase, located within a distance of less than  a few kpc, can soon be observed by Earth-based neutrino observatories~\cite{Odrzhywolek2004APh,Odrzhywolek2004AcPPB,Asakura_ApJ818}. Consequently, recent efforts have focused on calculating precise neutrino spectra and luminosities from pre-supernova stellar~cores.

In stellar matter, neutrino emission arises from various thermal processes and nuclear weak-interaction reactions.  Thermal (anti)neutrinos from pre-supernova cores are primarily produced via electron--positron pair annihilation (PA), with~their spectra
fully determined by the matter temperature  ($T$), density ($\rho$), and~electron fraction ($Y_e$)~\cite{Kato2015ApJ}.

During the final stages preceding collapse, as~the star undergoes silicon burning into iron, weak-interaction reactions involving atomic nuclei become the dominant source of (anti)neutrino production~\cite{Asakura_ApJ818,Patton2017ApJ_1,Patton2017ApJ_2}. The~expected pre-supernova neutrino spectra and the corresponding detector signals  were derived in Refs.~\onlinecite{Patton2017ApJ_1,Patton2017ApJ_2}. Notably, the~authors applied the methodology described in Ref.~\onlinecite{Langanke_PRC64} to compute the nuclear contribution to the spectra. This approach approximates each nucleus at given temperature-density conditions as undergoing a single transition from a parent to a daughter state using an effective $Q$-value ($Q_\mathrm{eff}$) method. This $Q$-value is adjusted to reproduce the average neutrino energy predicted by shell-model calculations for nuclear weak processes in stellar environments~\cite{ADNDT79_Langanke}. As~first noted in Ref.~\onlinecite{Misch_PRC94}, the~$Q_\mathrm{eff}$ method may fail to reproduce key spectral features, particularly when thermally excited nuclear states dominate the weak~transitions.

At typical pre-supernova temperatures ($10^9$\,K or higher), too many nuclear excited states become thermally populated making a state-by-state analysis of individual contributions to stellar weak-interaction processes computationally infeasible. Shell-model calculations address this challenge by employing the Brink hypothesis and the 'back-resonances' method~\cite{Langanke2000NuPhA}. In~our recent publications~\onlinecite{Dzhioev_Particles6,Dzhioev_MNRAS527,Dzhioev_IJMPE33}, we demonstrated that the contributions of thermally excited states to (anti)neutrino production can be accurately evaluated using the thermal quasiparticle random-phase approximation (TQRPA). This framework provides a thermodynamically consistent treatment of both excitation and de-excitation processes in hot nuclei. Unlike alternative approaches, the~TQRPA does not rely on the Brink hypothesis and inherently satisfies the detailed balance condition. Additionally, the~method enables the treatment of neutral-current nuclear de-excitation  where a hot nucleus emits neutrino-antineutrino pairs. The~potential significance of this process for generating high-energy pre-supernova (anti)neutrinos was first highlighted in Refs.~\onlinecite{Misch_PRC94,Patton2017ApJ_2} and later confirmed by TQRPA calculations~\cite{Dzhioev_Particles6,Dzhioev_MNRAS527,Dzhioev_IJMPE33}.

The present work extends our previous studies, and~here we apply the TQRPA to compute  the temporal evolution of luminosities and spectra for a specific
pre-supernova model and compare the TQRPA results with those  predicted by the effective  $Q$-value~method.

\section{Neutrino Spectra Within the TQRPA Framework  and Pre-Supernova~Model}\label{formalism}

In stellar matter, every nucleus at a given temperature-density ($T,\,\rho$)  condition can emit neutrinos and antineutrinos through various weak-interaction processes. These include continuum electron/positron capture (EC/PC), $\beta^\pm$-decay, and~nuclear de-excitation (ND) accompanied by the emission of a $\nu\bar\nu$-pair. It is important to note that EC (PC) and $\beta^+$ ($\beta^-$)-decay produce electron (anti)neutrinos only, whereas nuclear de-excitation generates $\nu\bar\nu$-pairs of all three neutrino flavors. In~this regard, the~ND process resembles thermal processes. However, unlike thermal processes, the~ND mechanism produces identical spectra for all (anti)neutrino flavors~\cite{Dzhioev_Particles6}.

Under stellar conditions, the~aforementioned reactions are predominantly governed by Gamow--Teller (GT) transitions: charge-exchange GT$_+$ (GT$_-$) transitions drive  electron (positron) capture and $\beta^+$ ($\beta^-$)-decay,  while  charge-neutral GT$_0$ transitions mediate the  ND process.
{In a thermally excited nucleus, both positive  and negative energy GT transitions are possible: electron and positron captures can proceed through both  positive  and negative energy transitions, while only negative energy transitions contribute to $\beta^\pm$-decay and $\nu\bar\nu$-pair emission.
The explicit formulations for the neutrino spectra expressed in terms of GT strength functions are detailed in Refs.~\onlinecite{Dzhioev_Particles6,Dzhioev_MNRAS527}. These calculations involve integrating the corresponding phase-space factors over the transition energy.
For charge-exchange processes, the~phase-space factors additionally depend on the electron  and positron distributions. The~latter are given by the Fermi distributions with temperature $T$ and chemical potentials $\mu_{e^-}=-\mu_{e^+}$
that depend on the temperature, matter density, and~electron fraction.  Since no charged leptons are involved in nuclear de-excitation, this process is independent of matter~density.}

We compute the strength functions using the TQRPA framework, a~technique that extends the quasiparticle random-phase approximation to finite temperature through the superoperator formalism in Liouville space~\cite{Dzhioev_PhPN53_1}. Within~the TQRPA framework, the~Gamow--Teller (GT) strength function is expressed in terms of transition matrix elements from the thermal vacuum to eigenstates (thermal phonons) of the thermal Hamiltonian:
\begin{equation}\label{str_funct}
 S_{\mathrm{GT}_{\pm, 0}}(E,T) = \sum_i \mathcal{B}^{(\pm,0)}_{i}\delta(E-\omega_i \mp \Delta_{np}).
\end{equation}

Here, $\mathcal{B}^{(\pm,0)}_{i} = |\langle Q_i\|\mathrm{GT}_{\pm, 0}\|0(T)\rangle|^2$ is the transition strength to the $i$-th thermal phonon state of a hot nucleus and $E_i^{(\pm, 0)} = \omega_i \pm \Delta_{np}$ is the respective transition energy; $\Delta_{np}=0$ for charge-neutral transitions, while  for charge-exchange transitions $\Delta_{np}=\delta\lambda_{np} + \delta M_{np}$, where $\delta\lambda_{np} = \lambda_n-\lambda_p$ is the difference between neutron and proton chemical potentials in the nucleus, and~$\delta M_{np}=1.293$\,MeV is the neutron--proton mass splitting. Note that eigenvalues of the thermal Hamiltonian, $\omega_i$, can be either positive or negative.  Consequently, the~strength functions~\eqref{str_funct}  for upward ($E>0$) and downward $E<0$ satisfy the detailed balance principle:
\begin{equation}\label{DB}
   S_{\mathrm{GT}_{\mp,0}}(-E,T)= \mathrm e^{-(E\mp\Delta_{np})/kT}  S_{\mathrm{GT}_{\pm,0}}(E,T).
\end{equation}

This property ensures the thermodynamic consistency of the TQRPA framework.  Following Refs.~\onlinecite{Dzhioev_Particles6,Dzhioev_MNRAS527,Dzhioev_IJMPE33}, our current TQRPA calculations are self-consistent employing the SkM* parameterization of the Skyrme effective nucleon--nucleon~interaction.

As in Ref.~\onlinecite{Dzhioev_IJMPE33},  we utilize the stellar evolutionary web interface~\cite{mesaweb} from the MESA project~\cite{Paxton2011ApJS,Paxton2013ApJS,Paxton2015ApJS,Paxton2018ApJS,Paxton2019ApJS} to simulate the evolution of a progenitor star with an initial mass of \mbox{$M=14M_{\odot}$}. The~remaining input parameters for our pre-supernova model are provided in Ref.~\onlinecite{Dzhioev_IJMPE33}.
For each evolutionary time step, MESA outputs temperature, density, electron fraction and isotopic composition as a function of the mass coordinate $m$. These quantities are  subsequently used to derive (anti)neutrino spectra and luminosities from both nuclear weak reactions and pair~annihilation.

\begin{figure}[t]
\includegraphics[width=0.54\textwidth]{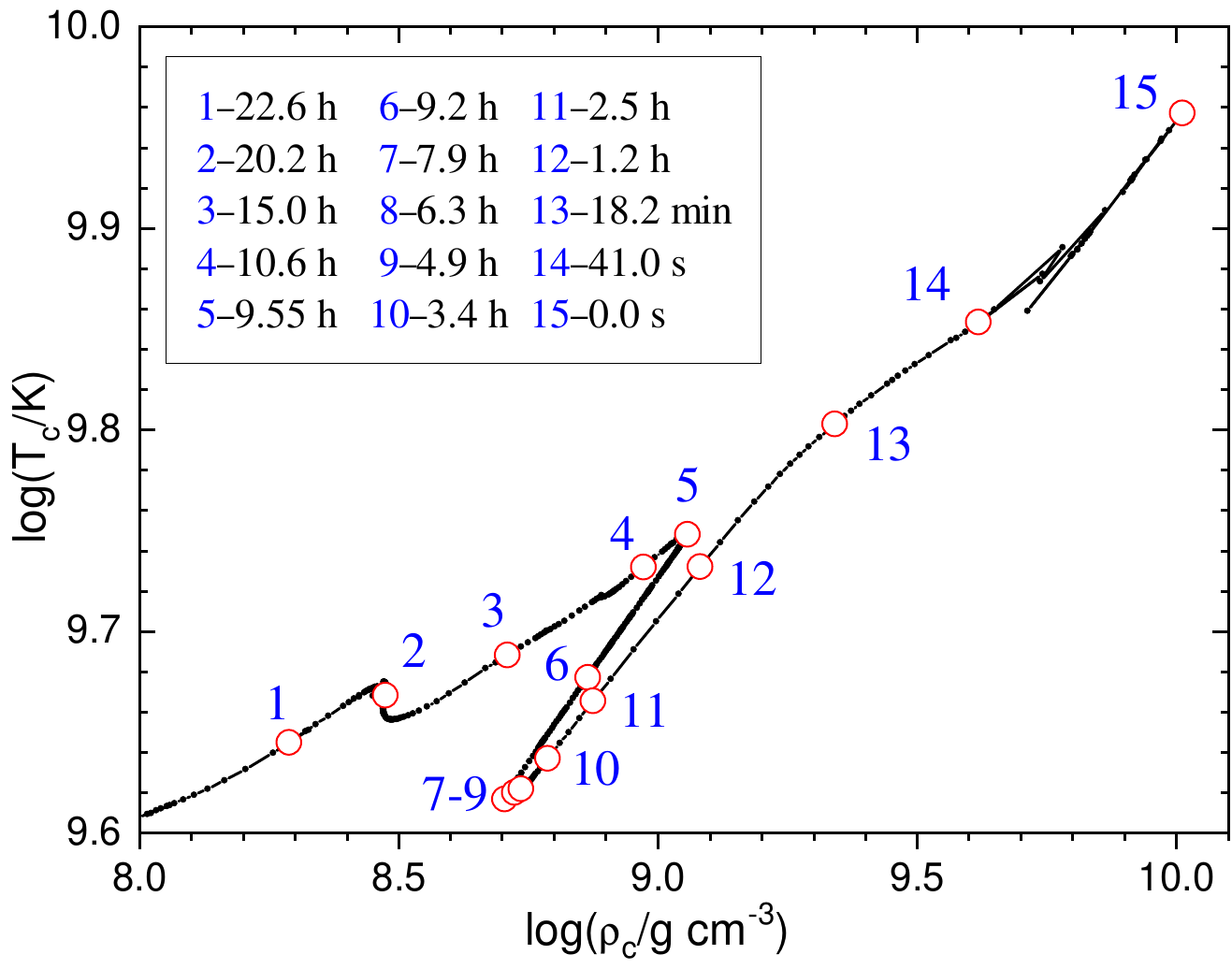}
\caption{The trajectory in the plane central temperature $T_c$ and central density $\rho_c$.  The~open circles correspond to the points at which the (anti)neutrino luminosities were calculated. For~each point we indicate the time~$t$ to~collapse.}
 \label{trajectory}
\end{figure}

Figure~\ref{trajectory} shows the evolution of the central temperature $T_c$ and density $\rho_c$ during the final day before core collapse ($t = 0$). The~plot reveals that both temperature and density increase nearly monotonically, except~during the period between  $t=9.55$\,h and  $t=7.9$\,h when $^{28}$Si burning (primarily into $^{54}$Fe) occurs in the layer surrounding the iron core. Throughout this entire time interval, $^{56}$Fe remains the dominant isotope in the core's hottest region, which extends up to approximately $m\approx1\,M_\odot$ (see Figure~2 in Ref.~\onlinecite{Dzhioev_IJMPE33}).
In addition to $^{56}$Fe, the~following isotopes were included
into calculations of the (anti)neutrino spectra and luminosities: $^{52,54}$Fe, $^{56}$Ni, $^{32}$S
and $^{28}$Si. Among~them, only $^{52}$Fe and $^{56}$Ni are unstable in their ground state and
produce $\nu_e$ through $\beta^+$-decay.

\section{Results}\label{results}

\subsection{Luminosities and Spectra of Pre-Supernova~Neutrinos}

\begin{figure}[t]
\includegraphics[width=0.55\textwidth]{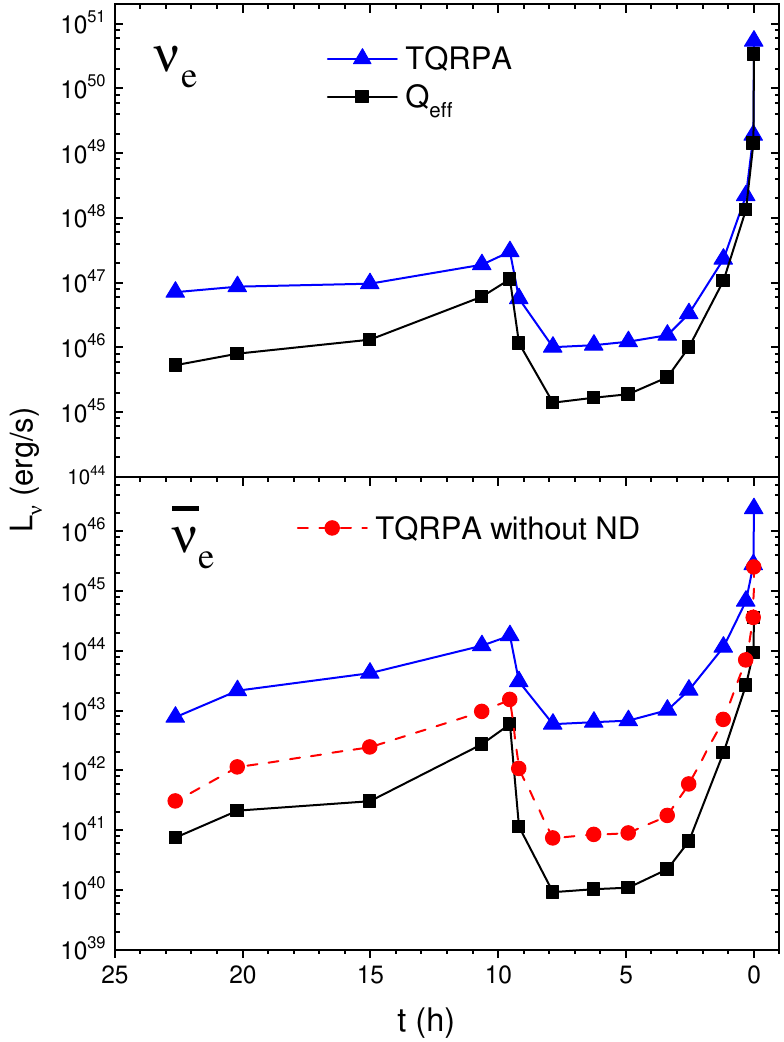}
\caption{Time evolution of the nuclear contributions to $\nu_e$ and $\bar\nu_e$ luminosities computed employing the TQRPA and $Q_\mathrm{eff}$ approaches. For~$\bar\nu_e$ we also show the luminosity without the ND~contribution.}
 \label{TQRPAvsLSSMnucl}
 \end{figure}

Let us first examine the time evolution of the considered pre-supernova model in terms of the (anti)neutrino energy luminosity. Initially, we focus on the luminosities of $\nu_e$ and $\bar\nu_e$  species from nuclear processes without including flavor oscillations.
In Figure~\ref{TQRPAvsLSSMnucl}, we compare the luminosities calculated using the TQRPA and $Q_\mathrm{eff}$ method.  First of all, we note that both methods predict a luminosity  peak at $t\approx 10$\,h followed by a minimum and a subsequent rapid increase. These variations clearly reflect changes in temperature and density along the stellar evolution trajectory (see Figure~\ref{trajectory}).

Although the temporal evolution of luminosities appears similar between the approaches, the~$Q_\mathrm{eff}$ method consistently predicts lower luminosity values than the TQRPA approach. {For~electron neutrinos, this discrepancy diminishes with increasing temperature and density, becoming negligible at the core collapse onset ($t=0$). This convergence can be understood by considering that electron capture is the main source of $\nu_e$~\cite{Dzhioev_Particles6,Dzhioev_MNRAS527,Dzhioev_IJMPE33}. At~$t=0$, under~high-temperature and high-density conditions  the electron chemical potential is large enough ($\mu_{e^-}\approx 8\text{--}9$\,MeV) and  EC on iron-group nuclei occurs primarily through GT$_+$ resonance excitation.
As both approaches result in roughly equal energies and transition strengths to the resonance state, the~luminosity calculations naturally converge at the point of core~collapse.}

At lower temperatures and densities, the~available electron energy is insufficient to excite the  GT$_+$ resonance. Under~these conditions, electron capture  occurs primarily through downward (negative energy) transitions from thermally excited states. The~significance of downward transitions was demonstrated in Ref.~\onlinecite{Dzhioev_IJMPE33} through comparative analysis of $\nu_e$ luminosities from hot versus cold nuclei.
Namely, it was shown that during the final day before core collapse downward transitions may amplify  the $\nu_e$ luminosity by over an order of magnitude. As~discussed in the Introduction, the~$Q_\mathrm{eff}$ method employs average neutrino energies derived from prior shell-model calculations. In~Ref.~\onlinecite{Dzhioev_MNRAS527}, through a direct comparison of TQRPA and shell-model results, we demonstrated that the thermodynamically consistent treatment of negative energy transitions in the TQRPA framework produces not only higher electron capture rates but also  greater neutrino energies.
This fundamental difference explains why the $Q_\mathrm{eff}$ method systematically predicts lower $\nu_e$ luminosities in the pre-collapse phase compared to the~TQRPA.

Regarding electron antineutrino  emission in nuclear processes, the~dominant mechanism is charge-neutral nuclear de-excitation. As~shown in the lower panel of Figure~\ref{TQRPAvsLSSMnucl}, the~ND process enhances the $\bar\nu_e$ luminosity by approximately an order of magnitude. Since charge-neutral de-excitation is not taken into account  in shell-model calculations, the~$Q_\mathrm{eff}$ method significantly underestimates these luminosities.
 It should be mentioned that  for nuclear processes, emission of heavy-lepton flavor (anti)neutrinos  occurs exclusively through the ND process, and~the corresponding luminosities and spectra coincide with that of electron (anti)neutrinos.

{Let us now examine how the inclusion of thermal processes influences luminosity predictions in both approaches.   Recall that the dominant thermal process is pair annihilation. As~demonstrated in Refs.~\onlinecite{Dzhioev_MNRAS527,Dzhioev_IJMPE33}, the~PA process contributes to $\nu_e$ emission significantly less than electron capture. Consequently, $\nu_e$ luminosities remain virtually unchanged when including thermal processes and both approaches yield results consistent with the upper panel of Figure~\ref{TQRPAvsLSSMnucl}.}

\begin{figure}[t]
\includegraphics[width=0.55\textwidth]{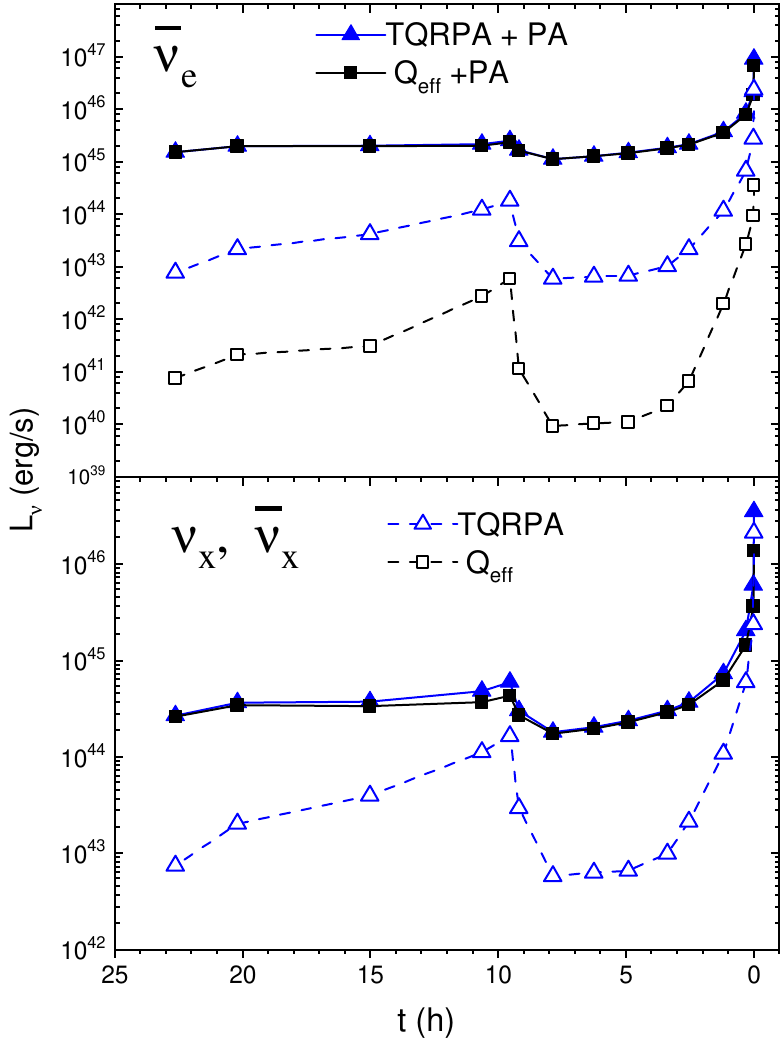}
\caption{Time evolution of the total $\bar\nu_e$ and $\nu_{\mu,\,\tau}$,  $\bar\nu_{\mu,\,\tau}$ luminosities computed employing the TQRPA and $Q_\mathrm{eff}$ approaches. For~comparison we also show (empty symbols) the luminosities from   nuclear processes. Note that within the $Q_\mathrm{eff}$ method $\nu_{\mu,\,\tau}$,  $\bar\nu_{\mu,\,\tau}$ can be produced only in thermal~processes.}
 \label{TQRPAvsLSSMtotal}
\end{figure}

The situation is different for other (anti)neutrino species, $\bar\nu_e$ and $\nu_x$, $\bar\nu_x$ ($x=\mu,\,\tau$). {As~shown in Figure~\ref{TQRPAvsLSSMtotal}, their emission  primarily arises from pair annihilation.  For~$\bar\nu_e$ the PA contribution is more significant because $\bar\nu_e$ production via pair annihilation occurs through both neutral and charged currents~\cite{Misiaszek2006PhRvD}. Consequently, the~total $\bar\nu_e$ luminosities are in close agreement between both approaches throughout the entire time period. The~$\nu_x,\,\bar\nu_x$ luminosities also show consistent agreement between both approaches with discrepancies emerging only at peak temperatures  when the  ND and PA luminosities become comparable.
Thus,  at~$t\approx 0,\,10$\,h the TQRPA predicts slightly greater $\nu_x,\,\bar\nu_x$ luminosities than the $Q_\mathrm{eff}$~method.}

\begin{figure}[t]
\includegraphics[width=0.9\textwidth]{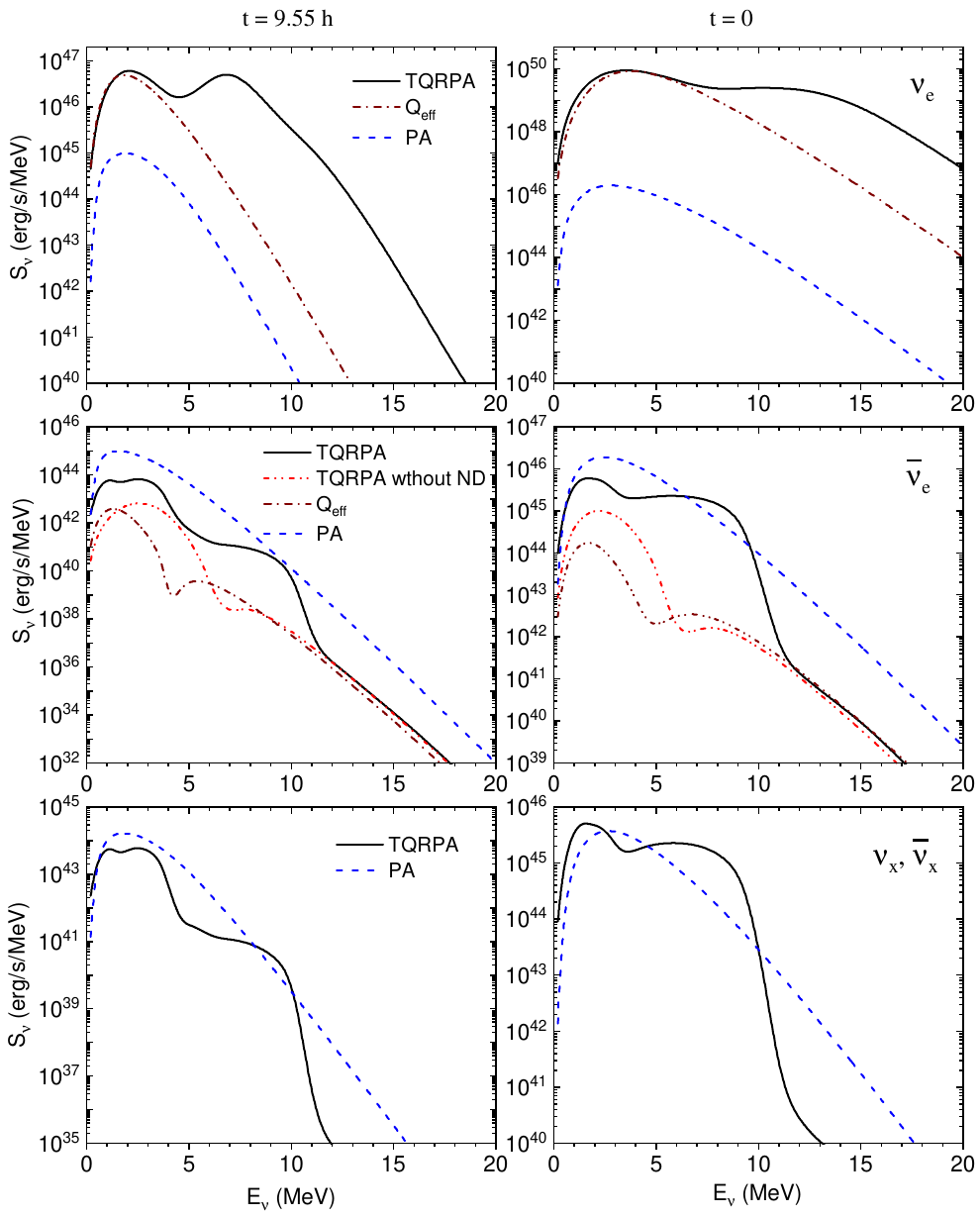}
\caption{Comparison of (anti)neutrino spectra from nuclear and thermal process at $t = 9.55$\,h
(left panels) and $t = 0$ (right panels). For~$\nu_e$ and $\bar\nu_e$  we also compare the TQRPA and $Q_\textrm{eff}$ spectra from nuclear processes. }
 \label{SpectraNoOscill}
\end{figure}

Figure~\ref{SpectraNoOscill} shows the energy luminosity spectra at selected times $t=0$ and $t=9.55$\,h with contributions from nuclear and thermal processes displayed separately. Both approaches predict a low-energy peak in the $\nu_e$ spectra originating from electron capture via GT$_+$ resonance excitation. The~TQRPA spectra additionally exhibit a distinct high-energy peak (absent in $Q_\mathrm{eff}$ results) caused by negative energy transitions from thermally excited states. Thus, including negative energy transitions not only increases $\nu_e$  luminosity but also hardens the~spectrum.

As shown in the middle panel of Figure~\ref{SpectraNoOscill}, without~the ND component the TQRPA and  $Q_\mathrm{eff}$  spectra of $\bar\nu_e$  display a double-peaked structure originating from positron capture and $\beta^-$-decay. Both approaches indicate that without nuclear de-excitation, pre-supernova $\bar\nu_e$ spectra is dominated by pair annihilation. However, the~TQRPA reveals that including ND processes  significantly enhances the impact of nuclear reactions on $\bar\nu_e$   production. Notably, in~the $7\text{--}9$\,MeV energy range, the~high-energy ND component becomes comparable to or even surpasses the PA contribution. As~discussed in Refs.~\onlinecite{Dzhioev_Particles6,Dzhioev_MNRAS527}, this high-energy component stems from the de-excitation of the GT$_0$ resonance. The~relative contribution of the ND process is even more significant for heavy-lepton flavor (anti)neutrinos (see the lower panel of Figure~\ref{SpectraNoOscill}), as~their PA contributions are intrinsically~smaller.

\subsection{Neutrino~Oscillations}

Thus far, we have examined (anti)neutrino emission processes in the pre-supernova phase without considering neutrino oscillations. However, the~flavor composition of the pre-supernova neutrino flux at Earth differs from that at production due to flavor oscillations. In~this analysis, we focus specifically on electron (anti)neutrinos, as~these are most relevant for detection. When accounting for both vacuum oscillations and the Mikheyev--Smirnov--Wolfenstein effect, the $\nu_e$  flux at Earth can be expressed as~\cite{Kato2015ApJ}:
\begin{equation}\label{F_oscill}
  F_{\nu_e} = p F^0_{\nu_e}+(1-p) F^0_{\nu_x}.
\end{equation}

Here $F$ represents either the spectral or luminosity flux, and~the superscript $0$ denotes the unoscillated flux~(For brevity,
in~\eqref{F_oscill} we omit the geometric factor due to the distance to the star). {An~analogous expression holds for $\bar\nu_e$ with the replacement $\nu\to\bar\nu$ and $p\to\bar p$. The~energy independent quantities $p$ and $\bar p$ are the survival probabilities. They are connected with the neutrino mixing angles as $p=\sin^2\theta_{13}\approx0.024$, $\bar p=\cos^2\theta_{12}\cos^2\theta_{13}\approx0.676$} for the normal hierarchy (NH) of neutrino masses and  $p=\sin^2\theta_{12}\cos^2\theta_{13}\approx0.300$, $\bar p=\sin^2\theta_{13}\approx0.024$ for the inverted hierarchy (IH) (see e.g.,~Ref.~\onlinecite{Kolupaeva2023}).

\begin{figure}[t]
\includegraphics[width=0.9\textwidth]{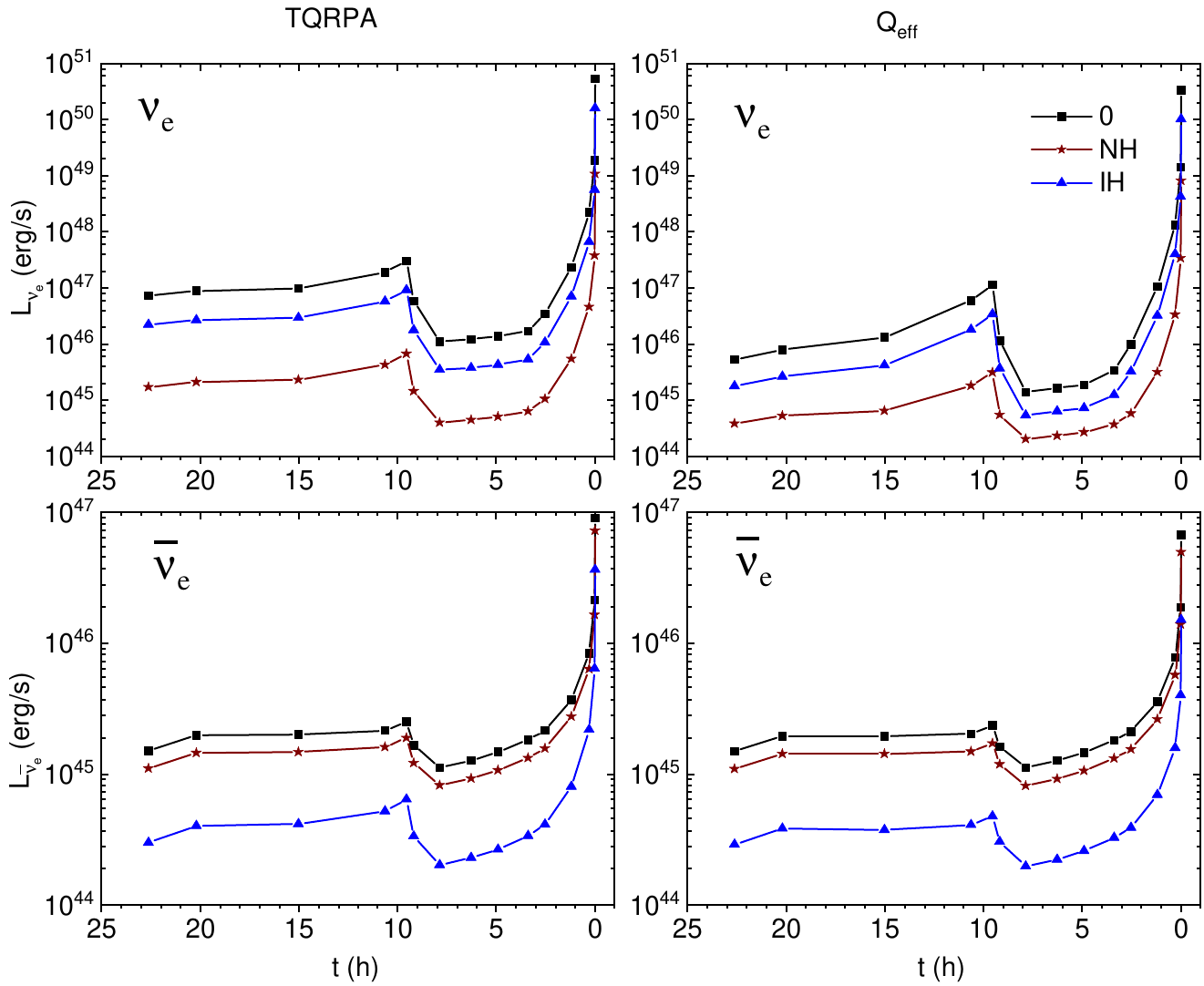}
\caption{Time evolution of the total $\nu_e$ (upper panels) and $\bar\nu_e$ (lower panels) luminosities. On~each
plot, the~luminosities obtained with the normal (NH) and inverted (IH) hierarchy are compared
with the original unoscillated luminosity (0). The~left and right plots show the results obtained employing the TQRPA and $Q_\mathrm{eff}$ approaches. }
 \label{lum_oscill}
 \end{figure}

In Figure~\ref{lum_oscill}, we compare the time evolution of unoscillated total  $\nu_e$ and $\bar\nu_e$ luminosities (computed using both the TQRPA and $Q_\mathrm{eff}$ approaches) with the corresponding luminosities after accounting for normal (NH) and inverted (IH) hierarchy oscillations.
 The figure clearly demonstrates that $\nu_e$ luminosity experiences strong suppression under the NH, while the suppression is less pronounced for the IH. Conversely, NH oscillations only marginally reduce the original $\bar\nu_e$ luminosity, whereas IH leads to substantially greater suppression. These effects can be directly understood from Equation~\eqref{F_oscill}. Specifically, the~survival probability $p\ll 1$ for the NH, and~since $L^0_{\nu_x}\ll L^0_{\nu_e}$, this results in a dramatic suppression of $L^\mathrm{NH}_{\nu_e}$. A~similar mechanism explains the stronger reduction of $\bar\nu_e$ luminosity under IH~oscillations.

As discussed in Ref.~\onlinecite{Dzhioev_IJMPE33}, inverted hierarchy  oscillations effectively replace nearly the entire original  $\bar\nu_e$ flux with $\bar\nu_x$, while no equivalent substitution occurs for $\nu_e$. This $\nu_e$  preservation amplifies the impact of nuclear temperature effects on oscillated $\nu_e$ luminosities. The~thermal effects on  oscillated $\nu_e$ luminosities become particularly apparent when comparing the TQRPA and $Q_\mathrm{eff}$  results. Following from our earlier examination of  Figure~\ref{TQRPAvsLSSMnucl}, the~substantial discrepancy between unoscillated $L^0_{\nu_e}$  values stems from the fact that the $Q_\mathrm{eff}$ method fails to capture downward GT$_+$ transitions from thermally excited states. For~the same reason, oscillated luminosities  $L^\mathrm{NH,IH}_{\nu_e}$  obtained by using the $Q_\mathrm{eff}$ method are generally suppressed with respect to the TQRPA results.
In contrast,  $\bar\nu_e$  luminosity after oscillations remains predominantly driven by pair annihilation. Here, thermal enhancement of nuclear processes produces only marginal increase in oscillated TQRPA luminosity compared to $Q_\mathrm{eff}$  predictions, except~during high-temperature phases at $t\approx 0\ \mbox{or}\ 10$\,h where nuclear de-excitation  contributions become~significant.

\begin{figure}[t]
\includegraphics[width=0.9\textwidth]{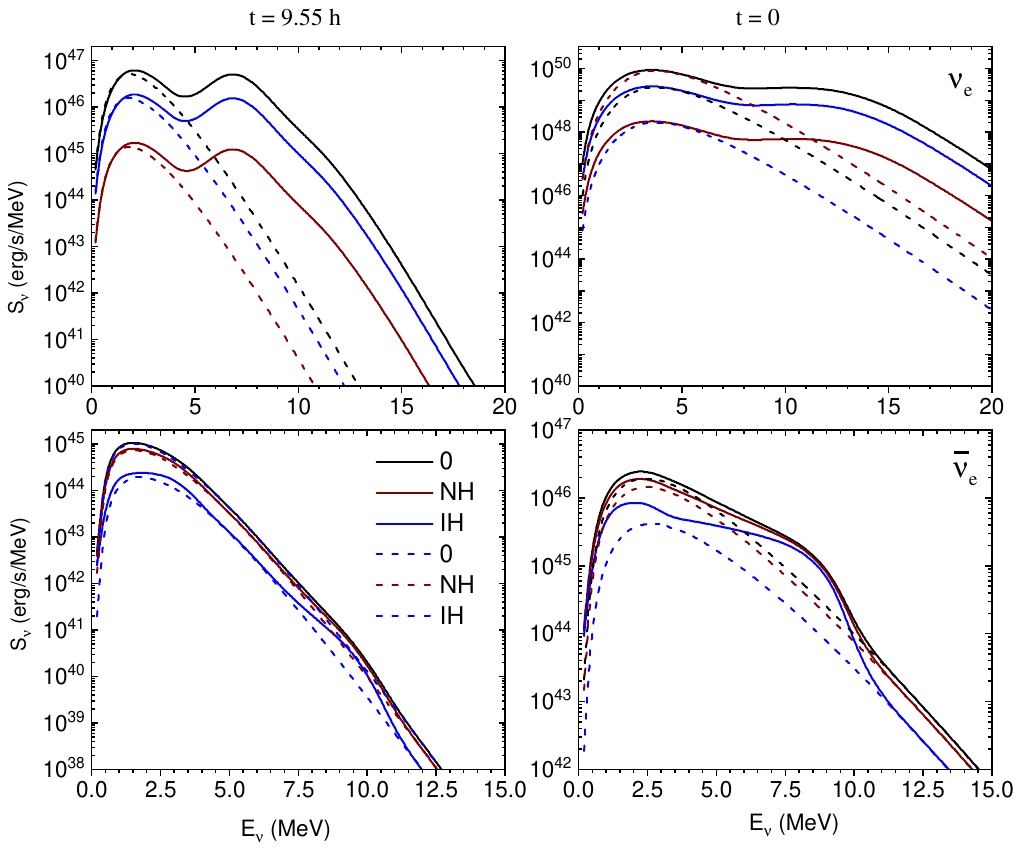}
\caption{The total spectra for $\nu_e$ (upper panels) and $\bar\nu_e$ (lower panels) obtained with the normal
(NH) and inverted (IH) hierarchy at $t = 9.55$\,h and $t = 0$ are compared with the original
unoscillated spectra (0). On~each plot we compare the spectra obtained  with the TQRPA (solid lines) and $Q_\mathrm{eff}$ method (dashed lines).}
 \label{spectra_oscill}
\end{figure}

Figure~\ref{spectra_oscill} illustrates the combined effects of flavor oscillations and nuclear thermal excitations on electron (anti)neutrino spectra calculated using both the TQRPA and $Q_\mathrm{eff}$ approaches at two time points $t=9.55$\,h and $t=0$. For~$\nu_e$ spectra, oscillations primarily reduce the overall flux magnitude while leaving the spectral shape largely unaffected. Both approaches yield nearly identical low-energy peaks,   the~TQRPA showing an additional high-energy peak from electron capture on thermally excited nuclear states. The~average $\nu_e$ energy remains stable due to incomplete $\nu_e\to\nu_x$ conversion.

For $\bar\nu_e$, oscillations enhance the thermal modifications of the TQRPA spectrum. This is particularly evident in the inverted hierarchy (IH) scenario where the original $\bar\nu_e$ flux is largely replaced by $\bar\nu_x$, leading to a strong suppression across most of the spectrum except for the energy range $E_\nu=7\text{--}10$\,MeV, where the ND contribution dominates. Since the ND spectrum is identical for all  flavors, it remains unchanged by oscillations. As~a result, the~interplay between the ND process and IH oscillations boosts the fraction of high-energy $\bar\nu_e$  in the spectrum, with~this effect being more pronounced at higher temperatures (i.e., at~ $t=0$). Regarding the spectrum computed within the $Q_\mathrm{eff}$ method, the~absence of the ND process prevents any increase in the average neutrino energy. For~the same reason, in~the inverted hierarchy scenario the spectrum experiences significantly stronger suppression compared to the TQRPA~approach.

\section{Conclusions}\label{conclusion}

In this work, we have investigated how nuclear temperature influences pre-supernova (anti)neutrino emission by comparing luminosities and spectra derived from two methods: the recently developed Thermal Quasiparticle Random-Phase Approximation (TQRPA) and the conventional effective $Q$ approach. Based on our earlier studies~\cite{Dzhioev_Particles6,Dzhioev_MNRAS527,Dzhioev_IJMPE33}, we demonstrated that the thermodynamically consistent treatment of Gamow--Teller (GT) transitions in hot nuclei within the TRQPA approach enhance the nuclear contribution to (anti)neutrino emission in comparison with the
$Q_\mathrm{eff}$ method results.  For~ $\nu_e$, downward GT$_+$ transitions from thermally excited states not only increase the high-energy luminosity but also produce a harder spectrum compared to the $Q_\mathrm{eff}$ method. While flavor oscillations suppress the $\nu_e$ luminosity in both approaches, they leave the spectral shape unchanged.  In~the case of  $\bar\nu_e$  emission at high temperatures, TQRPA calculations reveal that the neutral-current de-excitation process plays a more prominent role when oscillations are accounted for, significantly enhancing the fraction of high-energy $\bar\nu_e$ in the spectrum. Unlike the TQRPA approach, no such enhancement is observed in the $Q_\mathrm{eff}$ method. These results have critical implications for detecting pre-supernova (anti)neutrinos in terrestrial experiments. {\color{black} In Ref.~\onlinecite{Dzhioev_MNRAS527}, we demonstrated that the ND process can significantly enhance the detection rate of pre-supernova $\bar{\nu}_e$ via inverse beta-decay. However, as~illustrated in Figure~\ref{spectra_oscill}, during~the pre-supernova phase the $\nu_e$ flux exceeds the $\bar{\nu}_e$ flux by orders of magnitude. This suggests that detection channels sensitive to electron neutrinos may play a crucial role in observing the final hours of stellar evolution. In~future work, we plan to investigate how thermal effects---particularly the hardening of the $\nu_e$ spectrum predicted by the TQRPA framework---influence the detectability of pre-supernova $\nu_e$ through alternative interaction channels, including elastic scattering on electrons, charged- and neutral-current interactions with nuclei, and~coherent elastic neutrino–nucleus scattering.}




%

\end{document}